\documentclass[conference]{IEEEtran}
\IEEEoverridecommandlockouts
\pdfoutput=1
\usepackage{cite}
\usepackage{amsmath,amssymb,amsfonts}
\usepackage{algorithmic}
\usepackage{graphicx}
\usepackage{listings}
\usepackage{textcomp}
\usepackage{booktabs}
\usepackage[bookmarks=false]{hyperref}
\usepackage{todonotes}
\usepackage[binary-units=true]{siunitx}
\usepackage{float}
\usepackage{xcolor}
\def\BibTeX{{\rm B\kern-.05em{\sc i\kern-.025em b}\kern-.08em
    T\kern-.1667em\lower.7ex\hbox{E}\kern-.125emX}}
    
\IEEEoverridecommandlockouts\IEEEpubid{\makebox[\columnwidth]{978-1-7281-4973-8/20/\$31.00
~\copyright~2020 IEEE \hfill} \hspace{\columnsep}\makebox[\columnwidth]{ }}
\begin{document}

\title{A Resource Efficient Implementation of the RESTCONF Protocol for OpenWrt Systems}

\author{\IEEEauthorblockN{Malte Granderath}
\IEEEauthorblockA{Computer Science, Jacobs University Bremen, Germany\\
malte.granderath@gmail.com}
\and 
\IEEEauthorblockN{J\"urgen Sch\"onw\"alder}
\IEEEauthorblockA{Computer Science, Jacobs University Bremen, Germany\\
j.schoenwaelder@jacobs-university.de}
}

\maketitle

\begin{abstract}
In recent years, the open source operating system OpenWrt has become a popular option for replacing proprietary firmware on networking devices such as home routers or access points. In order to configure an OpenWrt system, like setting up firewall rules, the user has to either sign in to the web interface or use SSH to manually change configuration files on the device. While the current approach is sufficient for small home networks, it only allows for limited automation of management tasks and configuration management becomes time-consuming, for example, on larger campus networks where access control lists on OpenWrt access points need updates regularly.

This paper describes our efforts to implement the RESTCONF configuration management protocol standardized by the IETF on OpenWrt systems that have limited CPU and memory resources. We detail our design choices that make our implementation resource efficient for the use cases we target and we compare our implementation against other similar solutions. Our implementation is available on GitHub under an open source license\footnote{Available from https://github.com/mgranderath/orc under BSD-3}.
\end{abstract}


\section{Introduction}
The IETF has standardized a new network management protocol called RESTCONF \cite{RFC8040} that supports a subset of the functionality of NETCONF \cite{RFC6241}. RESTCONF uses HTTP \cite{RFC7230} as a substrate and hence any tool or programming language that can invoke HTTP operations can be used to implement management applications. RESTCONF provides a programmatic interface for installing, changing, or deleting configuration data, which is accessible via HTTP methods. The protocol supports different data encodings (e.g., XML \cite{RFC7950} and JSON \cite{RFC7951}). The exposed API is data-model-driven. The data models are written in the YANG \cite{RFC7950} data modeling language and define the resources exposed by the RESTCONF server. The original version of RESTCONF supported a single unified datastore abstraction exposing both configuration and state data. The more recent for network management datastore architecture \cite{RFC8342} has led to RESTCONF extensions that can expose multiple configuration datastores and a separate operational state datastore \cite{RFC8527}.

The OpenWrt open source project is creating and maintaining an embedded operating system based on the Linux kernel, which is mostly used on networking devices, such as home routers and access points. All OpenWrt components have been optimized to run efficiently on embedded devices with limited resources. A typical system running OpenWrt has at least 8MB of flash memory and at least 64 MiB of main memory. Systems with 16 MiB flash and 128 MiB main memory are quite common at the time of this writing. OpenWrt implements a fully writable and accessible file system and it provides a package management system allowing users to easily install additional applications \cite{openwrt_project}. The OpenWrt system implements a special approach to configuration file management, which is called the Unified Configuration Interface (UCI). It enables the unified management of all configuration information of the system. UCI keeps configuration information separated from the parts of the system that are essentially read-only.

While looking for a RESTCONF implementation that runs efficiently on OpenWrt platforms, we found that open source implementations were either attempts to port implementations originally written for much more resource rich platforms over to OpenWrt or they were written with embedded systems in mind but still designed to consume resources even when no management interactions take place. Hence, we wanted to find out how resource efficient a RESTCONF implementation running under OpenWrt can be made. In particular, we wanted to avoid long running processes that constantly consume memory. Since we also wanted to achieve good response times, it was crucial to minimize the startup time of the RESTCONF subsystem.

The rest of the paper is structured as follows. In the Section \ref{sec:uci}, we provide a brief introduction to the UCI subsystem used by OpenWrt to store configuration information. We then review a few basics of the RESTCONF protocol and the YANG data modeling language in Section \ref{sec:restconf} before we discuss related work in Section \ref{sec:related}. We describe our solution in Section \ref{sec:solution} and provide an evaluation in Section \ref{sec:evaluation} before we conclude the paper in Section \ref{sec:conclusions}.

\section{Unified Configuration Interface}
\label{sec:uci}

Linux systems usually store configuration files in the \texttt{/etc} directory tree and the configuration files use various file formats. OpenWrt tries to unify the format and location of the configuration files through the unified configuration interface (UCI). UCI configuration files are stored in the \texttt{/etc/config} directory and all UCI configuration files have the same format. Due to this unifying approach, configuration data can be exposed and manipulated easily via a common application programming interface (API).

A UCI file usually represents a UCI package, i.e., it contains all configuration information for a specific component of the system. UCI configuration files are divided into sections. Sections start with a \texttt{config} keyword followed by a section type name identifying the type of the section. Sections can be named or unnamed: a named section is identified by an additional name and the same section can appear multiple times with different names; unnamed sections usually only appear once in a configuration file. Inside the sections, configuration options are defined using the \texttt{option} keyword followed by the option type name and a value. The \texttt{list} keyword declares a list of name value pairs, utilizing the same list name for several items. Note that the format does not allow for nested sections.

\begin{figure}[htb]
\begin{lstlisting}
config system
    option hostname "OpenWrt"
    option timezone "UTC"
    # ...

config interface "en0"
    option ip6addr "2001:db8::42/64"
    option ip6gw   "2001:db8::1"
    # ...

config vnstat
    list interface "en0"
    list interface "en1"
    # ...
\end{lstlisting}
\caption{Examples of UCI configuration sections}
\label{fig:uci-example}
\end{figure}

Figure \ref{fig:uci-example} shows examples of UCI configuration sections. The first section is unnamed and defining system options. The second section is named and defines options for the interface named \texttt{en0}. This third section is again unnamed and defines a list of interfaces that should be used for statistics collection.

OpenWrt ships with a small C library (\texttt{libuci}) to access UCI files. It exposes multiple functions that can be used to read, write, or modify the configuration data. The library handles the locking of the configuration files during changes and prevents write collisions. It also exposes some further additional functions that help validate values and names of sections. The \texttt{uci} command line utility can be used interactively on the command line to make changes. The web user interface \texttt{LuCI} is written in Lua \cite{IFC18} and uses the \texttt{libuci} as well to access and manipulate configuration files.

\section{RESTCONF and YANG}
\label{sec:restconf}

YANG \cite{RFC7950} is a data modeling language that was designed to model data accessed and manipulated by configuration management protocols like RESTCONF or NETCONF. YANG models configuration and state data in a hierarchical fashion with essentially four elements:

\begin{itemize}
    \item A \textit{leaf} schema node models simple data of a simple data type. A leaf node instance has exactly one value and no child nodes.
    \item A \textit{leaf-list} schema node models an ordered sequence of simple data of one specified simple data type. A leaf-list instance can have zero, one, or multiple values.
    \item A \textit{container} schema node groups related schema nodes into a sub-tree. A container instance does not have a value but it can contain child nodes of any type.
    \item A \textit{list} schema node defines a sequence of list entries. Each list entry acts as a container. A list schema node for configuration data is required to define the child elements that form a key identifying list instances.
\end{itemize}

YANG has been used to define data models covering, for example, network interfaces \cite{RFC8343} or general system aspects \cite{RFC7317}.

The RESTCONF specification maps hierarchical models defined in YANG modules to a hierarchical set of resources that can be accessed and manipulated by invoking HTTP methods on them. The data exchanged between a RESTCONF client and a RESTCONF server can be encoded in either XML or JSON. RESTCONF uses HTTP methods to realize the create, read, update and delete (CRUD) operations needed to manipulate configuration data \cite{RFC8040}:

\begin{itemize}
    \item The \texttt{OPTIONS} method is used to discover which operations are possible for a specific resource.
    \item The \texttt{GET} method is used to retrieve data and header fields of a specific resource.
    \item The \texttt{HEAD} method only returns the header fields of a \texttt{GET} request.
    \item The \texttt{POST} method is used to create a new data resource (or to invoke an operation)
    \item The \texttt{PUT} method can create new data resources, like POST, but, if the resource already exists, it replaces the existing resource.
    \item The \texttt{PATCH} method is used to modify an existing resource.
\end{itemize}

\section{Related Work}
\label{sec:related}

Several of open-source RESTCONF implementations can be found by searching through public git repositories and digital libraries.

JetConf is a RESTCONF implementation that is implemented entirely in Python. It supports the JSON encoding and only HTTP/2 as the HTTP transport. JetConf has dependencies on several non-standard Python libraries that have to be installed. This means that only systems having the Python interpreter and those libraries installed will be able to run the RESTCONF implementation \cite{jetconfarchitecture}. The developers mention that this implementation is not suitable for low-end OpenWrt routers \cite{CZNICRESTCONF}. Note that OpenWrt systems usually do not come with Python installed since the Python language environment is relatively expensive in terms of resource requirements.

MD-SAL is an infrastructure component of the OpenDaylight (ODL) project that supports NETCONF and RESTCONF. It is implemented in Java and supports several transports and payload formats \cite{MDSAL}. RESTCONF is part of the MD-SAL artifact and cannot be installed standalone. There is only limited Java Virtual Machine (JVM) support on OpenWrt and only stripped down JVM's are available for installation such as JamVM \cite{jamvm}. Hence, this implementation is not suitable for OpenWrt devices.

Clixon is a YANG based configuration manager with both NETCONF and RESTCONF interfaces and an interactive command line tool. Clixon provides the core system and can be used as-provided but it also exposes an API for plugins, such that additional functionality can be implemented. It also has dependencies on some non standard libraries. Clixon is built as a system of services that utilize Remote Procedure Calls (RPC) to communicate with each other. When running RESTCONF there have to be three processes running. It needs a web server with support for FastCGI, the backend daemon and the restconf daemon. Therefore, on systems that only have small numbers of configuration changes, resources are consumed whether or not RESTCONF is needed at that point in time \cite{clixon}.

\section{Solution Overview}
\label{sec:solution}

For our RESTCONF implementation, \textit{orc}, to be able to run on all devices that run OpenWrt, we focused on adhering to the functionality that is provided in a base installation of OpenWrt. We assume that the devices do not have to handle hundreds or thousands of RESTCONF requests per minute but rather have to handle configuration changes that arrive only occasionally. This means that our implementation should not consume any resources when no configuration transactions are taking place. For integration with OpenWrt the UCI system should be used as a datastore. This means that we have to map between the different data formats. An additional requirement that we set for our implementation is to depend on as few non-standard libraries as possible and to utilize only libraries that have a small package size.

\subsection{Architecture and Design Decisions}

\begin{figure}[htb]
\centering
\includegraphics[width=\columnwidth]{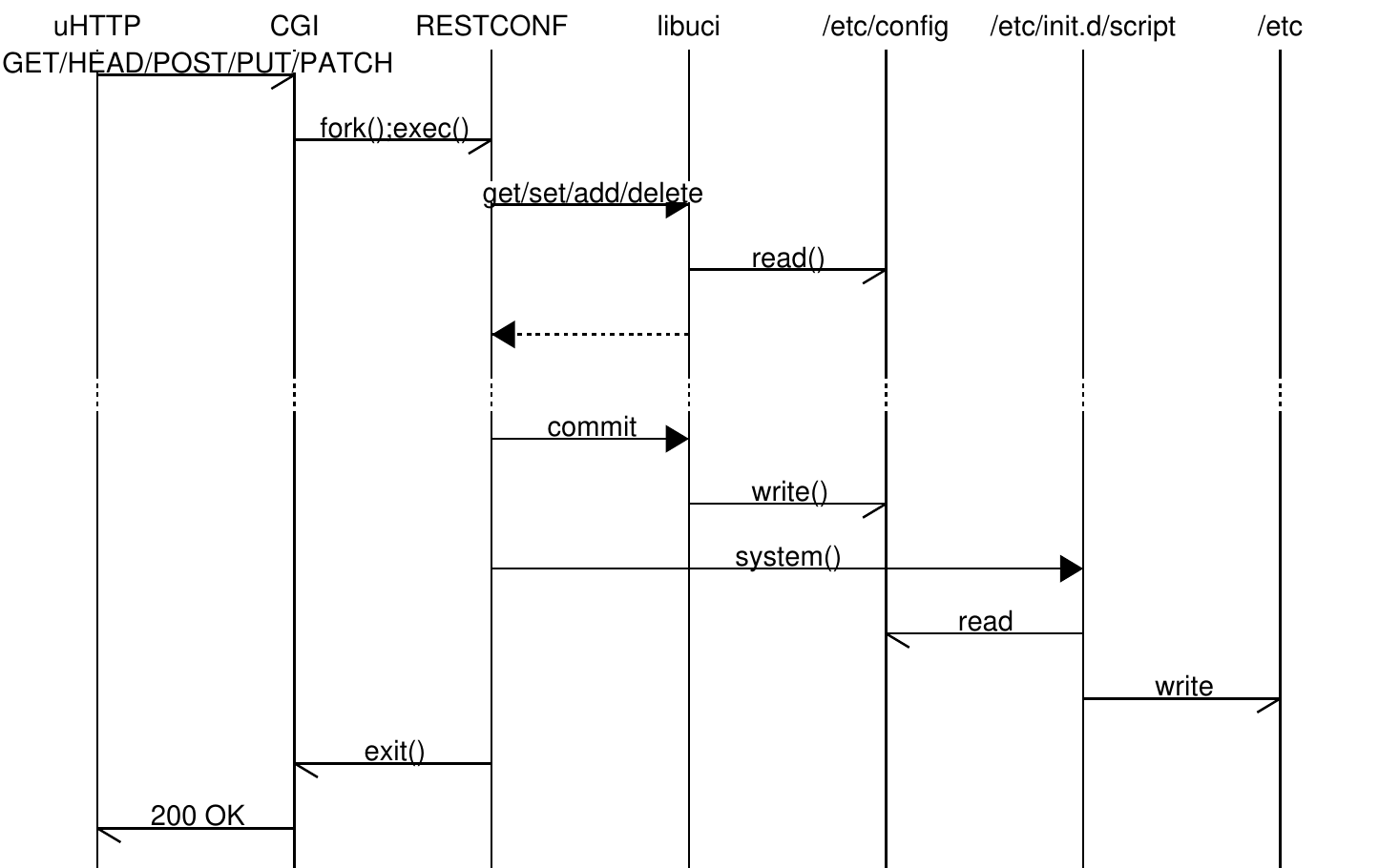}
\caption{Architecture of the implementation}
\label{fig:implementation_arch}
\end{figure}

The architecture of our implementation is best explained using the time sequence diagram shown in Figure \ref{fig:implementation_arch}.

The \textit{uHTTP} web server was implemented specifically for the OpenWrt operating system and it supports the Common Gateway Interface (CGI) \cite{RFC3875}. We decided to build our implementation \textit{orc} as native binary that can be executed via the CGI mechanism of \textit{uHTTP}, see Figure~\ref{fig:implementation_arch}. This ensures that other applications can use the same web server (most notably LuCI) and \textit{orc} will only be executed when requested. This reduces the consumption of resources compared to having our implementation built integrated with its own web server. This design has the added benefit that the native binary can be used with any web server that supports the CGI.

When \textit{orc} is executed, it interprets the request using the provided YANG models and then, depending on the request, reads or writes to the UCI using the \textit{libuci} library that is pre-installed on OpenWrt systems. Our implementation also maps between the data formats so that JSON configuration data received from a RESTCONF client can be stored in the proper UCI format. The \textit{libuci} writes and reads the configurations stored in the \texttt{/etc/config} directory and then, if required, a \texttt{/etc/init.d} script can be executed to convert the UCI configuration into formats that are interpreted by the applications and services. This is visualized by the \texttt{/etc} directory in Figure \ref{fig:implementation_arch}.

\subsection{Mapping YANG and UCI}
\label{sec:uci_yang_extensions}

For utilizing the UCI system we need to be able to interpret the UCI file format in relation to YANG data models. This can be achieved by adding annotations to the YANG modules. The following annotations can be added to YANG modules:

\begin{itemize}
    \item The \texttt{package} annotation is used to set the UCI package name of the sub-tree.
    \item The \texttt{section} annotation is used to set the UCI section type of the sub-tree. This has to be specified for a list.
    \item The \texttt{section-name} is used to set the UCI section name of the sub-tree.
    \item The \texttt{option} annotation is used to set the UCI option and list names for leafs and leaf-lists.
    \item The \texttt{leaf-as-name} annotation is optional and can be used to define which leaf holds the value that should be used as \texttt{section-name}
\end{itemize}

\begin{figure}[htb]
    \centering
    \begin{lstlisting}
module example {
  uci:package "example";
  container device {
    uci:section "device";
    uci:section-name "device";
    leaf enabled {
      uci:option "enabled";
      ...
    }
    list interface {
      uci:section "interface";
      uci:leaf-as-name "name";
      ...
    }
  }
}
    \end{lstlisting}
    \caption{Example of an annotated YANG module}
    \label{fig:annotated_yang_example}
\end{figure}

The following rules apply for these YANG extensions:

\begin{enumerate}
    \item The \texttt{package}, \texttt{section} and \texttt{section-name} annotations can be overridden in the same sub-tree by specifying them again. This allows for splitting up a complex module into different UCI files and sections.
    \item The \texttt{option} annotation can only be used to annotate leafs or leaf-lists and cannot be overridden.
    \item If just a \texttt{section} is defined for a container, then \texttt{section-name} is implicitly declared as empty.
    \item A list node must have a \texttt{section} defined but no \texttt{section-name} (as seen in Figure \ref{fig:annotated_yang_example}). It can, however, use the \texttt{leaf-as-name} annotation to use a value of a leaf as a \texttt{section-name}.
    \item The UCI restrictions concerning allowed characters etc.~apply.
\end{enumerate}

\subsection{Representing a YANG model in JSON format}
\label{sec:conversion}

YANG models can be represented in the native YANG format or an XML rendering called YIN \cite{RFC7950}. We looked at several libraries for parsing the native YANG format or generic XML libraries for parsing the YIN format but they all come with significant overhead. We therefore decided to render YANG models into a JSON representation in an additional preprocessing step so that we can read YANG modules at runtime using a JSON parser. On OpenWrt, the \texttt{json-c} library comes preinstalled since it is used by other applications on the system.

\begin{figure}[htb]
    \centering
    \begin{lstlisting}
{
  "type": "module",
  "package": "example",
  "map": {
    "device": {
      "type": "container",
      "section": "device",
      "section-name": "device",
      "map": {
        "enabled": {
          "type": "leaf",
          "option": "enabled",
          ...
        },
        "interfaces" {
          "type": "leaf-list",
          "section": "interface",
          "leaf-as-name": "name",
          ...
        }
      }
    }
  }
}
    \end{lstlisting}
    \caption{The annotated YANG example converted to JIN}
    \label{fig:yang_to_json}
\end{figure}

Since there was no previous work that allowed for conversion of a YANG model into a JSON representation, a new conversion was developed. The first step is the conversion from YANG to YIN and the second step is the conversion of the YIN XML representation to JSON. The JSON is then processed to simplify it for our targeted usage. The YANG representation in JSON will be referred to as JIN from here on. Figure \ref{fig:yang_to_json} shows the conversion of the YANG module example shown in Figure \ref{fig:annotated_yang_example} into the JIN format. We have left out the mapping of the namespaces for improved readability.

\subsection{Converting UCI to JSON}

For returning JSON data in a response to a \texttt{GET} request, the existing UCI content has to be mapped to JSON based on the YANG data model. From the request URI that is passed to the CGI script, the JIN module can be traversed to find the targeted node. At each step of traversal, the JIN node is checked for UCI annotations, defined in Section \ref{sec:uci_yang_extensions}, and the UCI path is stored in an object that is passed on to the next iteration. Once the target node is reached the UCI path up to that node will be defined.

From this node, a recursive depth-first traversal is started that reads and stores the UCI annotations at each node and passes it on to the children. There are different cases depending on which node is reached during the traversal:

\begin{itemize}
    \item For \textit{leaf} or \textit{leaf-list}s, the UCI path should be complete pointing at a specific UCI \texttt{option} or \texttt{list}. This value is then read using \texttt{libuci} and returned to the parent.
    \item There is a particular case for \textit{lists}. As stated in Section \ref{sec:uci}, sections can be accessed using an index and, therefore, the number of sections of a section type can be used to iterate through the list, and the index is passed down to the children on continuing the recursion on the sub-nodes.
\end{itemize}

\begin{figure}[htb]
    \centering
    \begin{lstlisting}
module: example
  +--rw device
     +--rw name?          string
     ...
     +--rw interfaces* [name]
     |  +--rw name      string
     |  +--rw enabled   boolean
     ...
     +--rw applications* string
    \end{lstlisting}
    \caption{Simplified YANG module tree representation.}
    \label{fig:restconf_example_module}
\end{figure}

Taking the module from Figure \ref{fig:restconf_example_module} as an example of this process, with the URI \verb!/data/example:device!, the following traversal will be executed:

\begin{enumerate}
    \item \texttt{name} $\rightarrow$ read UCI value
    \item { 
        \texttt{interfaces} $\rightarrow$ for $i$ in \textit{list\_length} (list\_length = 1)
        \begin{enumerate}
            \item \texttt{interfaces[i].name} $\rightarrow$ read UCI value
            \item \texttt{interfaces[i].enabled} $\rightarrow$ read UCI value
        \end{enumerate}
    }
    \item {
        \texttt{applications} $\rightarrow$ read UCI value
    }
\end{enumerate}

At every step of the traversal, the return value of the child will be combined with the key from JIN to produce JSON.

\subsection{Converting JSON to UCI}

While for converting UCI to JSON a traversal through JIN has to be executed, for converting JSON to UCI, a parallel traversal of JIN and JSON has to be done. Every object in the JSON representation has to be compared to the nodes in JIN. Similarly to the conversion of UCI as JSON, the request URI is also used to traverse to the target node. During this traversal, the individual nodes can be checked for existence depending on whether it is a PUT or POST request.

After traversing to the target node, the root key of the JSON should represent the JIN resource that is targeted or its parent. From this, a depth-first traversal can be run on the combined JSON and YANG tree. In case the JSON is not targeted at adding a list item, the content has to be flattened according to the UCI annotations and no knowledge about the existing file content has to be known except if they exist. The JSON key-value pairs will be converted to UCI path, value and type triples that can then be written after traversal. The return cases of this recursive depth-first traversal are as follows:

\begin{itemize}
    \item For a \textit{leaf}, a triple of the UCI path, JSON value and type "option" will be returned.
    \item For a \textit{leaf-list}, a list structure of triples of the UCI path, JSON value and type "list" will be returned.
    \item For each item in a \textit{list} the recursive traversal is continued but with the index passed in the UCI path object. The results are combined into a list structure and returned.
    \item For a \textit{container} the traversal is continued, but an additional triple is added of UCI path, no value and type "container". This is later used to create the named section.
\end{itemize}

After this step, the JSON has essentially been flattened into a list of UCI paths to values. The changes can then be written using \texttt{libuci} by iterating through the list.

\begin{figure}[htb]
    \centering
    \begin{lstlisting}
{
  "example:device": {
    "name": "Router_0",
    "interfaces": [{
      "name": "eth0",
      "enabled": true
    }],
    "applications": [
      "uhttpd",
      "luci"
     ]
  }
}
    \end{lstlisting}
    \caption{Example JSON request for the module in Figure \ref{fig:restconf_example_module}}
    \label{fig:example_json_request}
\end{figure}

For example, the request URI \verb!/data/! with the JSON content in Figure \ref{fig:example_json_request} will output the following list when processed together with the \textit{restconf-example} module:

\begin{footnotesize}
\begin{verbatim}
"example.device", container,
"example.device.name", option, "Router_0"
"example.device.@interfaces[0].name", option, "eth0"
"example.device.@interfaces[0].enabled", option, "true"
"example.device.applications", list, "uhttpd"
"example.device.applications", list, "luci"
\end{verbatim}
\end{footnotesize}

In case a list item is to be added to an already existing list the above principles still apply, but instead of initializing the list index at zero, it will be initialized with the list length.

\subsection{Verifying JSON against YANG constraints}

One of the features of YANG and RESTCONF is the verification of configuration data against the constraints defined in the data model. Our implementation does not verify the data when reading from UCI since it is assumed that if there are other changes to the configuration files, they will be verified by the entity or tool causing those changes. However, for writing JSON as UCI, the verification of a subset of the YANG constraints has been implemented. The general verification of the JSON structure is completed while traversing through the tree.  

During the conversion from YANG to JSON (see Section \ref{sec:conversion}) the \texttt{typedefs} and imported types are extracted into a separate list. This list simplifies the lookup of type information during verification. The verification of types is implemented as part of the process of writing JSON as UCI. Whenever a leaf or leaf-list is reached during traversal, the content in the JSON is verified according to the following steps:

\begin{enumerate}
    \item The type declaration is retrieved either directly from YANG or the list of \texttt{typedef}s and imported types.
    \item The content is compared to the lexical representation of the base built-in types defined in RFC 7951 \cite{RFC7951}.
    \item In case a \texttt{pattern} or \texttt{range} restriction is defined, the content is also verified against these restrictions.
\end{enumerate}

Verification of \texttt{unique} and \texttt{key} semantics of lists has also been implemented. For lists, the \texttt{key} and \texttt{unique} statement values are first extracted from the converted module and then a simple iterative comparison is done on all items. In the case that a list item is added, the value of that item has to be compared to the already existing values, so in that case, the existing items are first read from the UCI and added into a list that is then verified. This is similar to the verification of leaf-lists, where only single values have to be compared.

\subsection{Limitations}

The main restriction that is imposed by the difference in UCI and nested data structures is the nesting of a container or list in another list. The problem in the current implementation is to indicate which child item, of a list item, belongs to each list item. This can easily be demonstrated by the following example.

\begin{figure}[htb]
    \centering
    \begin{lstlisting}
config interface
    option name "eth0"
    ...
    
config interface
    option name "eth1"
    ...
    
config nested nested
    \end{lstlisting}
    \caption{Nested Container in List item.}
    \label{fig:limitation_example}
\end{figure}

If a container \texttt{nested} is added to the \texttt{interfaces} list in the \texttt{example} module (see Figure \ref{fig:restconf_example_module}), the UCI configuration file will look like shown in Figure \ref{fig:limitation_example}. The \texttt{nested} section can belong to either of the \texttt{interface} sections. This could be solved by changing the name of the section to contain a reference to the list item it belongs to.

\section{Evaluation}
\label{sec:evaluation}

The main goal of our implementation is to run with low overhead on systems with limited resources. We target deployment scenarios where management transactions occur infrequently. The following statistics were collected using a virtual machine (x86\_64 architecture) running OpenWrt 18.06.2 with \SI{64}{\mebi\byte} of RAM. Our implementation depends on several libraries. The package sizes of all dependencies can be found in Table \ref{tab:total_library_size}.

\begin{table}[htb]
    \caption{Total package library size}
    \centering
    \begin{tabular}{lc}
    \toprule
    \textbf{Library} & \textbf{Size}  \\
    \midrule
    libjson-c        & \SI{14}{\kibi\byte} \\
    libuci           & \SI{13}{\kibi\byte} \\
    libubox          & \SI{16}{\kibi\byte} \\
    \midrule
    Total            & \SI{43}{\kibi\byte} \\
    \bottomrule
    \end{tabular}
    \label{tab:total_library_size}
\end{table}

The memory space required by the libraries is kept at an absolute minimum, not even requiring \SI{0.05}{\mebi\byte} of the flash memory. In the official documentation, it is stated that a minimum of \SI{4}{\mebi\byte} of flash is needed for installation of OpenWrt and a minimum of \SI{32}{\mebi\byte} of RAM is recommended \cite{openwrt_supported_devices}.

\begin{table}[htb]
\caption{Evaluation result}
\centering
\begin{tabular}{lcccc}
\toprule
       & \multicolumn{2}{c}{\textbf{ORC}} & \multicolumn{2}{c}{\textbf{Clixon}} \\
\textbf{Test} & \textbf{Heap} & \textbf{Time} & \textbf{Heap} & \textbf{Time} \\
\midrule
GET    & \SI{1.1}{\mebi\byte} & \SI{0.00}{\second}  & \SI{5.9}{\mebi\byte} & \SI{0.09}{\second} \\
POST   & \SI{1.1}{\mebi\byte} & \SI{0.07}{\second}  & \SI{6.3}{\mebi\byte} & \SI{0.02}{\second} \\
PUT    & \SI{1.1}{\mebi\byte} & \SI{0.07}{\second}  & \SI{6.3}{\mebi\byte} & \SI{0.02}{\second} \\
DELETE & \SI{1.1}{\mebi\byte} & \SI{0.01}{\second}  & \SI{6.1}{\mebi\byte} & \SI{0.01}{\second} \\
\bottomrule
\end{tabular}
\label{tab:evaluation_result}
\end{table}

Our \textit{orc} implementation has been tested against the clixon implementation. We measured the maximum heap size and the time to complete requests. For testing we used an example YANG module that can be found on github\footnote{\url{https://github.com/mgranderath/orc/blob/master/yang/restconf-example.yang}}. The \texttt{time} utility was used to measure the time to completion and the \texttt{valgrind} massif tool was used to measure the maximum heap size.

Clixon needs a web server with support for FastCGI, the \texttt{clixon\_backend} service, and \texttt{clixon\_restconf} process to be constantly running. When utilizing the example setup from the clixon documentation, the following idle heap size can be collected: \texttt{clixon\_restconf} uses \SI{2.4}{\mebi\byte} and \texttt{clixon\_backend} uses \SI{2.9}{\mebi\byte}. The result of our measurements are summarized in Table \ref{tab:evaluation_result}. While our implementation has a lower maximum heap size, it has slower response times for \texttt{POST} and \texttt{PUT} requests. However, the response times are still well below \SI{0.1}{\second} and since \textit{orc} is not constantly consuming memory resources, we believe that \textit{orc} is a far more efficient solution for systems that do not receive a constant stream of configuration management requests.

\section{Conclusions}
\label{sec:conclusions}

We described the design of our RESTCONF protocol implementation for OpenWrt systems called \textit{orc}. We have outlined how \textit{orc} maps between the OpenWrt UCI format used as the configuration datastore and the nested data format used by YANG models.

Our comparison of \textit{orc} against an existing implementation (clixon) has shown that we can achieve a much lower memory footprint. Furthermore, \textit{orc} does not consume any resources when no RESTCONF transactions take place.

Our implementation has some limitations that arise from the mapping of the data models (flat UCI versus nested YANG). Our future work includes extending the JIN parser with more YANG features and reducing the limitations in the mapping of the data models. In addition, we are working towards integrating \textit{orc} with our implementation of standards for large-scale measurement frameworks and evolving standards for network and device security (denial of service signaling, remote attestation) that use RESTCONF as a substrate.

\section*{Acknowledgements}

This paper has received funding from the European Union’s Horizon 2020 Research and Innovation program under the CONCORDIA cybersecurity project (GA No. 830927).

\bibliographystyle{IEEEtran}
\begin{flushleft}
\bibliography{bibi}
\end{flushleft}

\end{document}